\documentclass[aps,prl,twocolumn,superscriptaddress]{revtex4} 

\begin{document}


\title{Spin singlet formation in MgTi$_2$O$_4$: evidence of a helical dimerization pattern}


\author{M. Schmidt}
\affiliation{ISIS Facility, Rutherford Appleton Laboratory, Chilton, Didcot, Oxon OX11 0QX, United Kingdom}
\author{W. Ratcliff II}
\affiliation{Department of Physics and Astronomy, Rutgers University, Piscataway, 
New Jersey 08854, USA}
\author{P. G. Radaelli}
\affiliation{ISIS Facility, Rutherford Appleton Laboratory, Chilton, Didcot, Oxon OX11 0QX, United Kingdom}
\affiliation{Department of Physics and Astronomy, University College London,  
London, WC1E 6BT, United Kingdom}
\author{K. Refson}
\affiliation{ISIS Facility, Rutherford Appleton Laboratory, Chilton, Didcot, Oxon OX11 0QX, United Kingdom}
\author{N. M. Harrison}
\affiliation{Department of Chemistry, Imperial College London, London, SW7 2AY, United Kingdom}
\affiliation{CCLRC, Daresbury Laboratory, Warrington, WA4 4AD, United Kingdom}
\author{S.W. Cheong}
\affiliation{Department of Physics and Astronomy, Rutgers University, Piscataway, 
New Jersey 08854, USA}


\date{\today}

\begin{abstract}
The transition metal spinel MgTi$_2$O$_4$ undergoes a metal-insulator
transition on cooling below $T_{M-I} = 260$ K. A sharp reduction of
the magnetic susceptibility below $T_{M-I}$ suggests the onset of a
magnetic singlet state.  Using high-resolution synchrotron and neutron
powder diffraction, we have solved the low-temperature crystal
structure of MgTi$_2$O$_4$, which is found to contain dimers with
short Ti-Ti distances (the locations of the spin singlets)
alternating with long bonds to form helices.  Band structure
calculations based on hybrid exchange density functional theory show
that, at low temperatures, MgTi$_2$O$_4$ is an orbitally ordered band
insulator.
\end{abstract}

\pacs{}

\maketitle

The interest in geometrically frustrated systems can be traced back to
the work of Linus Pauling \cite{pauling35} on the frozen disorder of
crystalline ice.  Frustration arises when geometrical constraints
promote a locally degenerate ground state.  In a periodic system with
this local geometry, there exists a manifold of degenerate ground
states, which may freeze on cooling forming ice or remain liquid down
to the lowest temperatures due to quantum effects.  A third
possibility is that of a phase transition that lowers the local
symmetry and lifts the degeneracy.  In 1956, Anderson considered the
ordering of charges or Ising spins on the B- site network of the
spinel structure~\cite{anderson56}. Spinels have the general formula AB$_2$X$_4$, where A
and B are metals and X is an anion.  The B site forms a network of
corner-sharing tetrahera, also known as the pyrochlore lattice, which
is geometrically frustrated. It can be shown that Pauling's
`ice rules' are equivalent to antiferromagnetic coupling between the
spins or to nearest-neighbor Coulomb repulsion between equal charges.
Anderson concluded that the spinels should have large low-temperature
residual magnetic or configurational entropy, similar to ice.  Also, 
he interpreted the Verwey transition of magnetite
(Fe$_3$O$_4$, a half-filled mixed-valence spinel) \cite{anderson56} as
an example of degeneracy-lifting transition.  Almost perfect
realizations of the `spin ice' concept were found much later in
rare-earth pyrochlores such as Ho$_2$Ti$_2$O$_7$ \cite{bramwell01} 
One might in principle ask what would happen if the `entities' (spins, charges,
etc.) at the nodes of the pyrochlore lattice had the tendency to form
pairs.  It is well known, for example, that early transition metals in
edge- or face-sharing octahedral coordination display strong
cation-cation interaction, which often leads to dimerization and spin
pairing \cite{goodenough60}.  The classic example of this behavior is
VO$_2$ (rutile structure, V$^{4+}$, $3d^1$, $S=1/2$), which undergoes
a metal-insulator transition at 340 K, associated with a structural
transition from the high-temperature tetragonal structure to a
monoclinic structure containing dimers with short V-V distances
(2.65\AA ) \cite{andersson56}.  
The magnetic susceptibility shows Curie-Weiss behavior above
T$_c$ and a nearly constant van Vleck-like contribution below T$_c$,
which is due to the formation of spin singlets associated with the V-V dimers.
The rutile structure has a strong 1-dimensional character, due to the
presence of chains of edge-sharing octahedra running along the
tetragonal $c$-axis.  Therefore, the
transition in VO$_2$ can be described as a classic Peierls transition,
associated with the splitting of the $d$ band into bonding and
antibonding branches \cite{wentzcovitch94}.  

In addition to early transition metals, low-spin cations with higher
$t_{2g}$ orbital occupancy can also form dimers.  Recently, we have
solved the low-temperature structure of the mixed-valent CuIr$_2$S$_4$
spinel \cite{radaelli02}, which displays simultaneous M-I,
paramagnetic-diamagnetic and structural transitions at 230 K.  In
CuIr$_2$S$_4$, iridium has an average oxidation state of $3.5+$, and
half of the Ir ions were found to form short metal-metal bonds in the
low-temperature insulating phase, giving rise to simultaneous charge
ordering and dimerization.  MgTi$_2$O$_4$ is another spinel with
closely related phenomenology.  In this compound, Ti is in the $3+$
oxidation state with the $3d^1$ electronic configuration and $S=1/2$,
which makes it a strong candidate for the observation of dimerization in a
single-valence system.  In
fact, MgTi$_2$O$_4$ undergoes a M-I transition on cooling below $T_{M-I}=260$ K,
accompanied by a strong decrease of the magnetic susceptibility and a
transition to a tetragonal structure \cite{isobe02}.  In this letter, we report the
solution and refinement of the low-temperature crystal structure of
MgTi$_2$O$_4$, as obtained from synchrotron x-ray and neutron powder
diffraction data, and the results of electronic structure calculations 
based on the structural data.  The structure was found to contain dimers with
short Ti-Ti distances (2.85\AA), the locations of the spin singlets.
The dimer ordering leads to the formation of helical chains of short
and long bonds running along the tetragonal $c$ axis, a unique example
of a chiral ordering of the spinel structure.  The electronic structure
is consistent with the opening of a 1 eV gap and the absence of
magnetic moments, and enables one to interpret the crystal structure
in terms of orbital ordering.

MgTi$_2$O$_4$ powder samples were prepared by solid-state reaction of
an MgO, TiO$_2$ and elemental Ti powder mixture sealed under vacuum in
a silica tube. The reactants were sintered at 1080$^\circ$C several
times, followed by grinding in a glove box to increase
homogeneity. Measurements of electrical resistivity and magnetic
susceptibility of our sample produced results similar to those
reported by Isobe and Ueda \cite{isobe02}. Synchrotron x-ray powder
diffraction patterns of MgTi$_2$O$_4$ were collected on beamline ID31
at the European Synchrotron Radiation Facility in Grenoble, France,
using a multi-analyzer, parallel beam geometry at a wavelength of
0.5\AA. The sample was sealed in a 0.5mm glass capillary and cooled
down to 275K (above the transition) and 200K (below the transition)
using a He blower. Medium- and high- resolution neutron powder
diffraction patterns were collected at the Rutherford Appleton
Laboratory in Chilton, UK using the GEM and HRPD instruments
respectively. High-resolution patterns were collected at 200K and
275K, while medium-resolution data were collected in the 20--275K
temperature range in 20K steps.  For the neutron experiments, 
MgTi$_2$O$_4$  was sealed in a 6mm vanadium can under Ar
atmosphere and cooled by means of a closed cycle refrigerator.  The
neutron sample resulted from the second synthesis batch, and was purer
than the x-ray sample.

Comparison of the 275K and 200K x-ray diffraction patterns confirmed
the cubic-tetragonal phase transition and enabled us to identify most
of the impurity peaks (mainly Ti$_2$O$_3$ and MgO ) present in the
pattern at the background level. A portion of the 200K x-ray
diffraction pattern is presented in Fig. 1 (lower panel). Apart from
the tetragonal splitting of the main Bragg peaks, a series of weak
superlattice reflections also appear in the pattern.  These can all be
indexed on the $\frac{a_c}{\sqrt{2}} \times  \frac{a_c}{\sqrt{2}}\times a_c$
primitive tetragonal unit cell, having half the volume of the cubic
spinel cell.  The extinction conditions uniquely identify a pair of
enantiomorphic, non-centrosymmetric space groups: {\em P4$_1$2$_1$2}
and {\em P4$_3$2$_1$2}, which are equivalent except for their
chirality.  The atomic coordinates in the new space group can be
freely refined based on the x-ray data, yielding a structural model
that was further refined based on the neutron powder diffraction data.
The final structural refinement was based on the combined GEM and HRPD
data sets (a portion of the latter is shown in the inset to Fig. 1),
and the results at 200 K (tetragonal phase) and 275 K (cubic phase)
are presented in Table I.  Inclusion of the x-ray data slightly
degrades the quality of the Rietveld fit, because of selective
broadening of the superlattice reflections in the tetragonal phase;
however, the calculated intensities of both main peaks and satellites
are in excellent agreement with the x-ray results (see Fig. 1).

The tetragonal structure of MgTi$_2$O$_4$, as derived from the
 refinement of the 200K data is shown in Fig. 2.  The structure
 contains only one Ti site, ruling out the possibility of charge
 disproportionation.  However, the center of symmetry at the Ti site
 is lost, so that Ti moves away from the center of the TiO$_6$ octahedron,
 and the six nearest-neighbor Ti-Ti distances become inequivalent.
 Two out six Ti-Ti bonds  ($s=2.849(7)$\AA\ and $l=3.152(7)$\AA)
 differ substantially from the Ti-Ti distance found in the cubic
 MgTi$_2$O$_4$ ($3.00362(1)$\AA).  The shortest distance is comparable
 to the close-contact distance in Ti metal ($2.896$\AA\ at room
 temperature), suggesting the formation of a metal-metal bond.  It
 should be noted that the  intra-dimer distance in VO$_2$ is
 $2.654$\AA\ \cite{andersson56}, which is also comparable to the V-V
 distance in V metal ($2.61$\AA).  In the cubic spinel structure
 (Inset of Fig. 2), the TiO$_6$ octahedra form edge-sharing `ribbons',
 so that the Ti-Ti bonds run in straight lines along six equivalent
 $[110]_c$ directions.  In the tetragonal structure, both the short
 and the long bonds run along four: $[011]_c$, $[01\bar{1}]_c$,
 $[10\bar{1}]_c$, $[101]_c$ directions of the cubic structure
 ($[112]_t$ direction), alternating with one of the intermediate bonds
 ($i_1=3.002(5)$\AA) in the sequence {\em `s-i$_1$-l-i$_1$'}. Ti-Ti-bond lines
 running along the $[100]_t$ direction ($[110]_c$, $[1\bar{1}0]_c$) are
 entirely made up of the other type of intermediate bonds
 ($i_2=3.0099(3)$\AA).  Neither of the Ti-Ti-bond lines is perfectly
 straight, the Ti-Ti-Ti bond angle being $174.7(2)^\circ$ and
 $178.3(1)^\circ$   along the $[100]_t$ and $[112]_t$ directions,
 respectively.  Refinements of the temperature-dependent neutron data
 indicate that the phase transition is abrupt, with no
 coexistence region  between the two phase (Fig. 1, upper panel).
 This is reflected in the splitting of the Ti-Ti bond lengths, which
 is about 80\% of the full value at 250 K and is fully saturated
 below 200 K.  The metal-insulator transition in MgTi$_2$O$_4$  is
 very sensitive to impurities.  The magnetic
 susceptibility data for the Mg$_{2-x}$Ti$_{1+x}$O$_4$ solid solution
 indicates that  a partial replacement of the Ti$^{3+}$ with Ti$^{4+}$ ($3d^0$) and Mg$^{2+ }$
  breaks up the dimerization chains  \cite{hohl96,wechsler89,isawa94}.
  Ti$^{3+}$ undimerized due to impurities  most likely contribute to the Curie component of the
 MgTi$_2$O$_4$ magnetic susceptibility \cite{isobe02,hohl96}.

From the topological point of view, the most interesting aspect of the
MgTi$_2$O$_4$ low-temperature structure is the dimerization pattern of
the alternating short and long bonds.  Here, the chiral nature of the
space group is clearly revealed in the formation of {\em `s-l-s-l'}
`helices' running along the $c$-axis (Fig. 2).  With our choice of
space group, the helices are left handed, but the right-handed space
group {\em P4$_3$2$_1$2} is also an allowed solution.  Several authors
have pointed out the relevance of spin chirality for magnetism and
transport on a pyrochlore lattice \cite{ohgushi00,yoshii00,taguchi01}.
However, our observation of chirality in the structural sector of a
pyrochlore lattice is extremely unusual, and immediately raises two
issues.  Firstly, it is interesting to consider whether the chiral
dimerization pattern is in any way related to the geometrical
frustration of the pyrochlore lattice.  On this point, one should
notice that, once the system `decides' to dimerize, the topology of
the problem changes drastically: the relevant lattice is no longer the
pyrochlore lattice itself, but its `medial' (or bond center) lattice
at 1/6 filling (only one bond out of 6 is a dimer).  Each Ti atom can
only be involved in one dimer; therefore, one occupied dimer site
precludes the occupancy of the 10 nearest-neighbor bonds.  At such low
filling, this rule clearly leads to a highly degenerate ground state,
but, arguably, the system is no longer frustrated because the {\em
local} degeneracy is absent.  Secondly, the nature of the spin singlet
state needs to be further investigated.  To this effect, we have
carried out band structure calculations using the CRYSTAL
code~\cite{crystal}.  All electron, triple valence basis sets were
used in which three independent radial  functions are included for all
valence states~\cite{CRYSTALwww}.  Electronic exchange and correlation
are approximated using hybrid exchange density functional theory in
which a component of the non-local Fock exchange is  included with the
local exchange and correlation approximated within the generalised
gradient approximation  (GGA) resulting in the B3LYP
functional~\cite{becke93,lee88}. This functional is currently very widely
used in molecular studies as it yields ground state energetics
significantly more reliably than GGA functionals.  Here it is
preferred to GGA as the component of the Fock exchange cancels in part
the erroneous electronic self-interaction inherent in  GGA functionals
and thus the B3LYP approximation yields a reliable estimate of the
band gap in a wide range of materials, including strongly interacting
transition metal oxides~\cite{muscat01}.  We note that initial
calculations based on GGA description failed to reproduce the opening
of the  band gap in the tetragonal phase.
The computed projected densities of
states (DOS) are plotted in Fig. 3 for the cubic and tetragonal
structures; the latter was found to have a lower total energy than the
former, as observed experimentally.  In the cubic phase, the states
just below the Fermi energy $E_F$ form a highly dispersive, two-fold
degenerate band. This band is a common feature of many early
transition-metal spinels \cite{satpathy87}, and originates
mainly from $3d$-$t_{2g}$ titanium orbitals, with the $3d_{xy}$,
$3d_{xz}$ and $3d_{yz}$ participating equally in each energy level.
In MgTi$_2$O$_4$, the Fermi energy lies in a deep `trough', where this
band partially overlaps with more localized states at higher energy
(also of $t_{2g}$ origin), which accounts well for the poor metallic
properties at room temperature.  In the tetragonal, low temperature
phase the situation is radically different:  four new bands with
$3d_{xz}$ and $3d_{yz}$ contributions {\em only} are split off below
the main $t_{2g}$ band.  Complete filling of this bands with eight
electrons in the double cell results in a gap of about 1 eV, in good
agreement with the electrical conductivity \cite{isobe02}, while the
$3d_{xy}$ states are pushed to higher energy.  These results are
readily interpreted on the basis of the observed crystal structure, by
noting that the short Ti-Ti bonds have the same orientation as the
occupied $t_{2g}$ orbitals, alternating between the $3d_{yz}$
($[011]_c$, $[01\bar{1}]_c$), and $3d_{xz}$ ($[10\bar{1}]_c$,
$[101]_c$) directions on different sites.  This is consistent with
orbital ordering, and with the short bonds becoming the location of
zero-spin Hund-Mulliken (molecular-orbital-like) singlets.

In summary, we have determined the crystal and electronic structures
of the high- and low-temperature phases of MgTi$_2$O$_4$.  At low
temperatures, the Ti atoms dimerise forming an unusual chiral pattern.
This results in orbital ordering, and in the opening of a gap between
the fully occupied and unoccupied levels, in good agreement with the
transport and magnetic properties of MgTi$_2$O$_4$.
\begin{acknowledgments}
We would like to thank Daniel Khomskii for fruitful discussion of this work and
Fran{\c c}ois Fauth for his help with the ESRF synchrotron experiment.
This work was supported in part by US DOE Office of Science grant
DE-FG02- 97ER45651 and by NSF-DMR-0103858 grant. Reproduction of this article, with the customary
credit to the  source, is permitted.
\end{acknowledgments}


%
\section{Figure captions}
Fig.~1 {\bf Bottom:} The synchrotron x-ray diffraction pattern of MgTi$_2$O$_4$ at
background level taken at 200K, $\lambda=0.5$\AA\ (the
strongest line has an amplitude of 32000 counts). The top to bottom
rows of ticks  mark the positions of MgTi$_2$O$_4$, Ti$_2$O$_3$ and
MgO Bragg peaks respectively. The triangles mark the superlattice
reflections, the circles mark an unidentified
impurity. The inset shows a portion of the refined high-resolution
neutron diffraction pattern of MgTi$_2$O$_4$ at 200K. The top and
bottom rows of ticks mark the positions of MgTi$_2$O$_4$ and
Ti$_2$O$_3$ Bragg reflections respectively. The splitting of the cubic
peaks is apparent upon comparison to the spinel pattern taken at
275K. {\bf Top:} The tetragonal
lattice constants of MgTi$_2$O$_4$ and Ti-Ti bond lengths as functions
of temperature.  The solid lines are guides for eye.

Fig.~2 The Ti-Ti bond structure in tetragonal MgTi$_2$O$_4$ at
200K. The red and purple bonds represent the shortest (dimerized) and
the longest bonds in the MgTi$_2$O$_4$ structure respectively. The
dashed and solid blue bonds mark the intermediate $i_1=3.002(5)$\AA\ and
$i_2=3.0099(3)$\AA\ Ti-Ti distances respectively. A portion of the Ti
tetrahedral connectivity is also shown.  The inset shows a fragment of
the spinel structure in the same orientation, visualized using
cation-anion polyhedra.  One of the `helices' is outlined in yellow.  

Fig.~3 Electronic density of states (DOS) for cubic (top) and the
tetragonal (bottom) MgTi$_2$O$_4$. The contribution of Ti and O 
is indicated with different line styles (the Mg contribution to DOS in negligible in this energy range).
The vertical line marks the Fermi energy.
\begin{table}[t]
\caption{Lattice constants, fractional coordinates of atoms and bonds
in MgTi$_{2}$O$_4$ at 200K ({\em P4$_1$2$_1$2}) and 275K ({\em
Fd3m}). The parameters originate from the Rietveld refinement of high-
and medium- resolution neutron powder diffraction patterns.}
\begin{tabular}{l l c c c c}\hline\hline
\small
 200K  & & $x$& $y$& $z$& $U_{iso}$ [\AA$^2$] \\\hline
Ti &8b & -0.0089(5)  &   0.2499(9)  &  -0.1332(4)  &    0.0125(2)\\
Mg &4a & 0.7448(3)  & 0.7448(3) &  0   &    0.0073(2)\\
O(1) &8b & 0.4824(2) &  0.2468(3) &  0.1212(2)  &  0.0064(2) \\
O(2) &8b & 0.2405(3)  & 0.0257(2)  & 0.8824(2)  &  0.0035(2)\\\hline
\multicolumn{6}{l}{$a=6.01329(4)$\AA, $c=8.4703(1)$\AA,  $R_{wp}=0.0467$, $R_{p}=0.0637$}\\
\hline\hline
 275K  & & $x$& $y$& $z$& $U_{iso}$ [\AA$^2$] \\\hline
Ti &16d & 1/2 & 1/2& 1/2 & 0.0117(1)\\
Mg &8a & 1/8 & 1/8 & 1/8 &0.0069(1)\\
O &32e & 0.25920(2)  &  0.25920(2) &   0.25920(2)  &   0.00522(5) \\\hline
\multicolumn{6}{l}{$a=8.49552(3)$\AA,  $R_{wp}=0.0402$, $R_{p}=0.0267$} \\
\hline\hline
\multicolumn{2}{c}{Bond {[\AA]}} & \multicolumn{1}{c}{Cubic, 275\,K} & \multicolumn{3}{c}{Tetragonal, 200\,K}\\\hline
\multicolumn{2}{c}{Ti-Ti } & \multicolumn{1}{c}{3.00362(1) $\times6$} & \multicolumn{3}{c}{2.849(7) $\times1$ }\\
\multicolumn{2}{c}{ } & \multicolumn{1}{c}{} & \multicolumn{3}{c}{3.002(5) $\times2$, 3.0099(3) $\times2$}\\
\multicolumn{2}{c}{} & \multicolumn{1}{c}{} & \multicolumn{3}{c}{3.152(7) $\times1$}\\
\multicolumn{2}{c}{Ti-O } & \multicolumn{1}{c}{2.0487(2) $\times6$} & \multicolumn{3}{c}{2.028(5), 2.023(5), 2.021(5)}\\
\multicolumn{2}{c}{} & \multicolumn{1}{c}{} & \multicolumn{3}{c}{2.080(5), 2.135(4), 2.017(4)}\\
\hline\hline
\end{tabular}
\end{table}
%
\end{document}